\documentclass[graybox]{svmult}

\usepackage{type1cm}        
\usepackage{makeidx}         
\usepackage{graphicx}        
\usepackage{multicol}        
\usepackage[bottom]{footmisc}

\usepackage{newtxtext}       %
\usepackage[varvw]{newtxmath}       

\usepackage{acronym}

\usepackage[round]{natbib} 

\usepackage{epigraph} 
\acrodef{ai}[AI]{Artificial Intelligence}
\acrodef{si}[JHI]{Joint Hybrid Intelligence}
\acrodef{hmi}[HMI]{Human-Machine Interface}
\acrodef{jae}[JAE]{Joint Agent Engineering}
\acrodef{hsi}[HSI]{Human-Swarm Interaction}
\acrodef{loa}[LoA]{Levels of Automation}

\makeindex             

\begin{document}

		\title*{An Approach to Joint Hybrid Decision Making between Humans and Artificial Intelligence}
		\titlerunning{An Approach to Joint Hybrid Decision Making}

\author{Jonas D. Rockbach\\ Sven Fuchs  and\\ Maren Bennewitz}
\authorrunning{Rockbach, Fuchs, \& Bennewitz}
\institute{Jonas D. Rockbach (Corresponding author) \at Human-Machine Systems, Fraunhofer FKIE, Wachtberg, Germany \\ Humanoid Robots Lab, University of Bonn, Bonn, Germany \\ \email{jonas.rockbach@fkie.fraunhofer.de}
	\and Sven Fuchs \at Human-Machine Systems, Fraunhofer FKIE, Wachtberg, Germany \\ \email{sven.fuchs@fkie.fraunhofer.de}
\and Maren Bennewitz \at Humanoid Robots Lab, University of Bonn, Bonn, Germany \\ \email{maren@cs.uni-bonn.de}}
%
%
\maketitle


\abstract{Due to the progress in artificial intelligence, it is important to understand how capable artificial agents should be used when interacting with humans, since high level authority and responsibility often remain with the human agent. 
However, integrated frameworks are lacking that can account for heterogeneous agents and draw on different scientific fields, such as human-factors engineering and artificial intelligence.
Therefore, joint hybrid intelligence is described as a framework abstracting humans and artificial intelligence as decision making agents. 
A general definition of intelligence is provided on the basis of decision making competence being applicable to agents of different sorts.
This framework is used for proposing the interrelated design space of joint hybrid intelligence being aimed at integrating the heterogeneous capabilities of humans and artificial intelligence.
At the core of this design space lies joint agent engineering with the goal of integrating the design subspaces operator training, artificial intelligence engineering, and interface design via developing joint agent patterns.
The ``extended swarming'' approach to human-swarm interaction is discussed as an example of such a pattern. 
}

\section{Introduction}
\label{sec:intro}

\epigraph{\textit{`Thus an intelligent machine can be defined as a system that utilizes information, and processes it with high efficiency, so as to achieve a high intensity of appropriate selection. If it is to show really high intelligence, it must process a really large quantity of information, and the efficiency should be really high.''}}{\textit{Mechanisms of Intelligence}, Ross Ashby~\cite[p. 298]{Conant1981}}

Automation has become increasingly capable due to the progress in \ac{ai}.
While more processes are automatized based on \ac{ai}, traditional problems of human-automation interaction also become more important~\cite[]{Hollnagel2005}.
In complex use cases with high stakes, such as search-and-rescue, full autonomy of \ac{ai} is still hard to come by, and not always desired given ethical, legal or organizational constraints.
On the other hand, introducing \ac{ai} into the system is expected to increase total performance.
Therefore, \ac{ai} and human capabilities need to be integrated in a way so that the resulting system can deal as effectively and efficiently as possible with the task at hand in comparison to \ac{ai} and human acting isolated with less total performance.
If the integrated system's capabilities exceeds that of the individuals' acting alone, it is regarded as a joint agent itself acting as a unit to cope with the task demands.

While different labels and approaches for joining humans and \ac{ai} are being used, such as human-machine symbiosis~\cite[]{Licklider1960}, augmented cognition~\cite[]{Stanney2009}, cognitive systems engineering~\cite[]{Hollnagel2005}, human-machine cooperation~\cite[]{Flemisch2012b} and human-autonomy teaming~\cite[]{ONeill2020a}, we refer to the joining of human capabilities and \ac{ai} as \ac{si}.
In contrast to other labels, \ac{si} explicitly couples concepts of human-factors engineering and \ac{ai} under the umbrella concept of intelligent agents~\cite[]{Rockbach}, a concept also used in the transdisciplinary sciences of cybernetics and cognitive science.
\ac{si} refers to the methodical integration of heterogeneous agents, human and \ac{ai}, into a joint hybrid agent demonstrating superior performance compared to human and \ac{ai} acting isolated~\cite[]{Rockbach}.
In line with human-autonomy teaming~\cite[]{ONeill2020a}, the focus is on \ac{ai} demonstrating high competence in specific tasks.

So far, the approach of \ac{si} lacks a unified theoretical and practical basis.
Therefore, in this work we summarize the theoretical background considerations of intelligence and how these can be used for designing joint hybrid agents.
More specifically, Sec.~\ref{sec:int} defines intelligence as decision making competence so it can be applied to different types of agents.
While a theoretical measurement of decision making competence with little resource utilization $c$ is proposed, it is discussed that intelligence is hard to estimate in practice.
However, it is shown how the formalization is useful for theorizing about designing intelligence, including the augmentation of agent intelligence with other agents.
This theoretical basis is used in Sec.~\ref{sec:hi} to define \ac{si} and its design space in the form of the \ac{si} triad.
At the core of the \ac{si} design space lies \ac{jae} with the goal to integrate and constrain the different design subspaces human training, \ac{ai} engineering, and interface design.
Here, joint agent patterns are developed to weave the interdependent design subspaces together.
Sec.~\ref{sec:hsi} briefly discusses such a joint agent pattern for \ac{hsi}, called ``extended swarming''.
Finally, Sec.~\ref{sec:con} summarizes the considerations.

\section{Intelligence}
\label{sec:int}

\subsection{Intelligence as Competent Decision Making}

\subsubsection{Agent Theory}
\label{subsubsec:agent_theory}
A general definition of intelligence is provided by early work in cybernetics, a discipline that has been defined as the (comparative) study of control and communication in animals and machines~\cite[]{Ashby1961,Wiener1961}.
Cyberneticists were enthusiastic about the idea to understand control and communication in different agent types, both natural and artificial.
The cybernetician Ashby defined intelligence simply by stating "\textit{intelligent is, what intelligent does}''~\cite[p. 297]{Conant1981}; thus, intelligence is being a capable decision maker and acting accordingly.
In the following, this definition will be specified and contextualized in the light of a recent stream of cognitive science, 4E cognition (embodied, embedded, enactive, and extended)~\cite[]{Newen2018}, that are a modern descendant of early cybernetic ideas. 
The objective is to define intelligence in a way so it can be applied to any agent, i.e., any system that acts upon the world based on the state of the world~\cite[]{Russell2016}, such as humans and \ac{ai}.

\begin{definition}\label{def:agent}
An agent is a system that can make a decision and act accordingly on the basis of the state of the world.
\end{definition}

\begin{figure}
	\centering
	\includegraphics[width=0.7\linewidth]{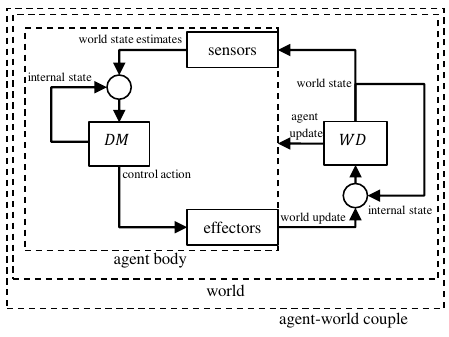}
	\caption{An agent body with sensors, decision matrix, and effectors is embedded in a particular world.}
	\label{fig:agent_world}
\end{figure}

An agent is a system embedded in a particular environment that can select actions based on its own capabilities which in turn are enabled by its body\footnote{This perspective includes software agents embedded in their virtual environments.} (Fig.~\ref{fig:agent_world})~\cite[]{Shapiro2019}.
According to enactivism, agent and world co-influence each other, resulting in an agent-world couple being defined by both the agent state and the world state~\cite[]{Varela1991a}.
Here, the agent actively acts on the world and is shaped by it, while the world can actively influence the agent.
The world state is a function of world dynamics $WD$ taking as inputs its previous state as well as influences from the agent.
The agent body on which basis actions are selected are usually grouped into three building blocks: sensors, decision making system, effectors.
The agent's decision matrix $DM$~\cite[]{Ashby1961}, modeling an agent's brain, computes control commands for the effectors based on its previous internal state as well as estimated world states provided by the sensors.
Note that an agent can only make a decision without acting.
Similarly, a change in world state is not a necessary condition to trigger a decision or action, since the agent could decide itself to perform some action based on its internal state.

\subsubsection{Agent Intelligence}
Now, given an agent placed in a world, it can be asked what makes it intelligent.
Based on Ashby's definition above, intelligence is defined as competent decision making, i.e., purposeful behavior attracting the agent-world system towards desired goal states~\cite[]{Rosenblueth1943}.
Intelligence is defined as follows:
\begin{definition}[Intelligence]
	\label{def:in_effec}
	Agent intelligence refers to the agent's decision making competence promoting the occurrence of goal states in the context of the world (effectiveness) while spending as few resources as possible (efficiency), as judged by an outside observer.
\end{definition}

\begin{figure}
	\centering
	\includegraphics[width=0.6\linewidth]{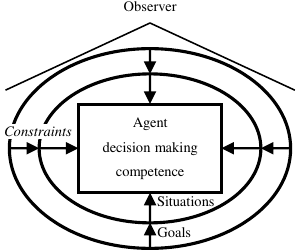}
	\caption{Intelligence depends on the observer constraints of agent purpose and agent-world situational contexts which constrain the required agent decision making competence.
	Adapted from ~\cite{Rockbach}.}
	\label{fig:intelligence}
\end{figure}

The definition is visualized in Fig.~\ref{fig:intelligence} and has three important properties.
First, intelligence is not binary, but rather a gradual. 
The better the agent is at making appropriate decisions (effectiveness) while using as few resources as possible (efficiency), the more intelligent it is.
Second, Def.~\ref{def:in_effec} takes observer dependencies into account~\cite[]{Brooks1999,Downing2015}.
Observer refers to the human evaluating the agent's intelligence~\cite[]{Maturana1970}.
Judging something as intelligent, and something as not intelligent, is primarily a matter of defining agent goals and the situational context of the agent-world system, and only secondary of actual agent properties.
These observer constraints provide the frame for the agent decision making competence; the combination of sensors, decision matrix, and effectors.

Thus, the constraints define the minimal required agent capabilities to solve the task.
If a task is not bound to resource expenditure, a random decision search process can be judged as satisfactory.
However, such a relaxed constraint is hardly ever the case, and even then an agent with less resource expenditure would be judged as more intelligent than a random search procedure.

Third, while the purpose of an agent always depends on the observer's judgment, a difference can be made for the situational context.
Given an agent-world system, there are variables which actually influence the agent's competence to reach goals, and variables the observer selected as relevant.
However, every system model is a simplification of a true phenomenon, if not wrong~\cite[]{Ashby1961}.
Therefore, the observer might evaluate an agent in situational contexts that are actually not influencing the agent's purposefulness or leave out relevant situational contexts.
While this distinction is an important one, we will not distinguish between the two in the following. 

Def.~\ref{def:in_effec} could be further extended by taking bodily limitations into account~\cite[]{Simon1982}.
For example, if an agent cannot sense a relevant variable of the agent-world couple because it does not have the appropriate sensors to do so, it can be seen as misleading to refer to the agent as being less intelligent.
However, some organisms are capable to extend or augment on their bodily limits~\cite[]{Hutchins1995}, such as apes using sticks, or humans developing scientific instruments to investigate phenomena not accessible by their biological bodies.
For this reason, the body is not taken into account in our definition of intelligence, which is aimed at defining ``output intelligence'' across organic and synthetic agents.

\subsection{Quantifying Agent Intelligence}

\subsection{Effectiveness: Probability to Reach Goal States}
\begin{figure}
	\centering
	\includegraphics[width=0.6\linewidth]{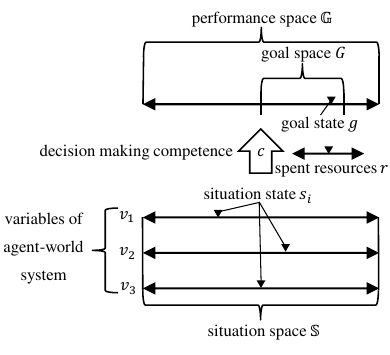}
	\caption{Visualization of the situation space $\mathbb{S}$ as a collection of agent-world system variable states $s$ impacting the performance space $\mathbb{G}$ via the agent's decision making competence $c$.
		$G$ as a subset of the performance space describes the tolerated performance values of the system, while $g$ is a particular state in the tolerated goal space range.
		The agent's decision making competence $c$ determines how effective and efficient the agent is at finding goal states $g$ in the context of the situation space $\mathbb{S}$.}
	\label{fig:gs}
\end{figure}

Quantifying agent intelligence can be split into considering effectiveness and efficiency. 
Based on Def.~\ref{def:in_effec}, an intelligent agent makes appropriate decisions to reach goal states in the context of possible situations.
In evaluating such decision making effectiveness of an agent, one observes the average agent's probability to reach goal states over possible situation states of the coupled agent-world system, 

\begin{equation}\label{eq:int_effective}
P(g|\mathbb{S})=\frac{\sum_{i=1}^{|\mathbb{S}|} P(g|s_i)}{|\mathbb{S}|}
\end{equation}

with $\mathbb{S}$ being the set of possible situations, also called the situation space, , $s \in \mathbb{S}$ being a particular system state $s_i=\{v_{1}^i,...v_{M}^i\}$ with $v$ being a selected system variable and $M$ being the length of the agent-world system model, and $|\mathbb{S}|$ being the size of the situation space as the number of all possible states $s$.
In turn, $G$ is the set of desired performance states out of the possible performance space $G \subseteq \mathbb{G}$, and $g \in G$ is a desired goal state. 
$|G|$ represents the goal space size as the number possible goal states.
The formalization is visualized in Fig.~\ref{fig:gs}.

In order to evaluate $P(g|\mathbb{S})$, the observer must at least select one performance metric $\mathbb{G}$ and the relevant agent-world system variables including their ranges to define the situation space $\mathbb{S}$ the agent could be confronted with.
$G$ is given by selecting a region of the performance space $\mathbb{G}$ that seems to represent purposeful behavior.
The thresholds of $G$ determine the difficulty of the competence estimation since a broad goal range makes it easier for the agent to reach a goal state. 
A narrow range tests for optimality, while a broad range tests for satisfactory competence~\cite[]{Simon1982}.

The average performance over different situation sets could be measured instead without defining an acceptable performance range, and one would have good reasons to do so.
Here the dichotomy of ``goal reached'' vs. ``goal not reached'' is used so the observer needs to make explicit what the purpose of the agent really is.
In addition, it is often relevant in practice to define acceptable performance ranges for agents since it provides a guideline for optimizing the agent's intelligence without undershooting (poor performance) or overshooting (wasting resources).
Taken together, $P(g|\mathbb{S})$ evaluates an agent's probability to find goal states, i.e., the demonstration of purposeful behavior, over different situational contexts.

\subsection{Efficiency: Spent Resources}
Considering two agents demonstrating the same effectiveness $P(g|\mathbb{S})$, the evaluation of efficiency, i.e., resources spent such as required time or energy, provides another quantitative distinction between them~\cite[]{Rockbach2022b}.
Intelligence is defined as effective decision making while spending as little resources as possible and is being calculated as decision making competence $c \in [0,1]$ by
\begin{equation}\label{eq:int_eff}
	c=rP(g|\mathbb{S})
\end{equation} 

with $r \in [0,1]$ being a normalized measurement of spent resources, $1$ being no resources spent, and $0$ being an undesired amount of resources spent. 
Fig.~\ref{fig:gs} summaries the formalization of intelligence $c$ as how capable the agent is at finding goal states $g$ while spending few resources $r$ in the context of possible situations $\mathbb{S}$.

\subsection{Estimating Decision Making Competence}
Evidently, $c$ is a theoretical construct and practically impossible to estimate in reality for complex problems since this would include the observation of a large part of the agent-world situation space multiple times. 
An estimation $\hat{c}$ can be achievable via statistical sampling for a selected subset of the situation space $S \subset \mathbb{S}$, which however relaxes the demand on agent intelligence, as will be discussed in the following.
However, it will be seen later that qualitative assessments based on $c$ can also be of use for human-\ac{ai} integration.

\subsection{Designing Agent Intelligence}

\subsubsection{Implications from Quantification}
\label{subsub:imp_in}

\begin{figure}
	\centering
	\includegraphics[width=1\linewidth]{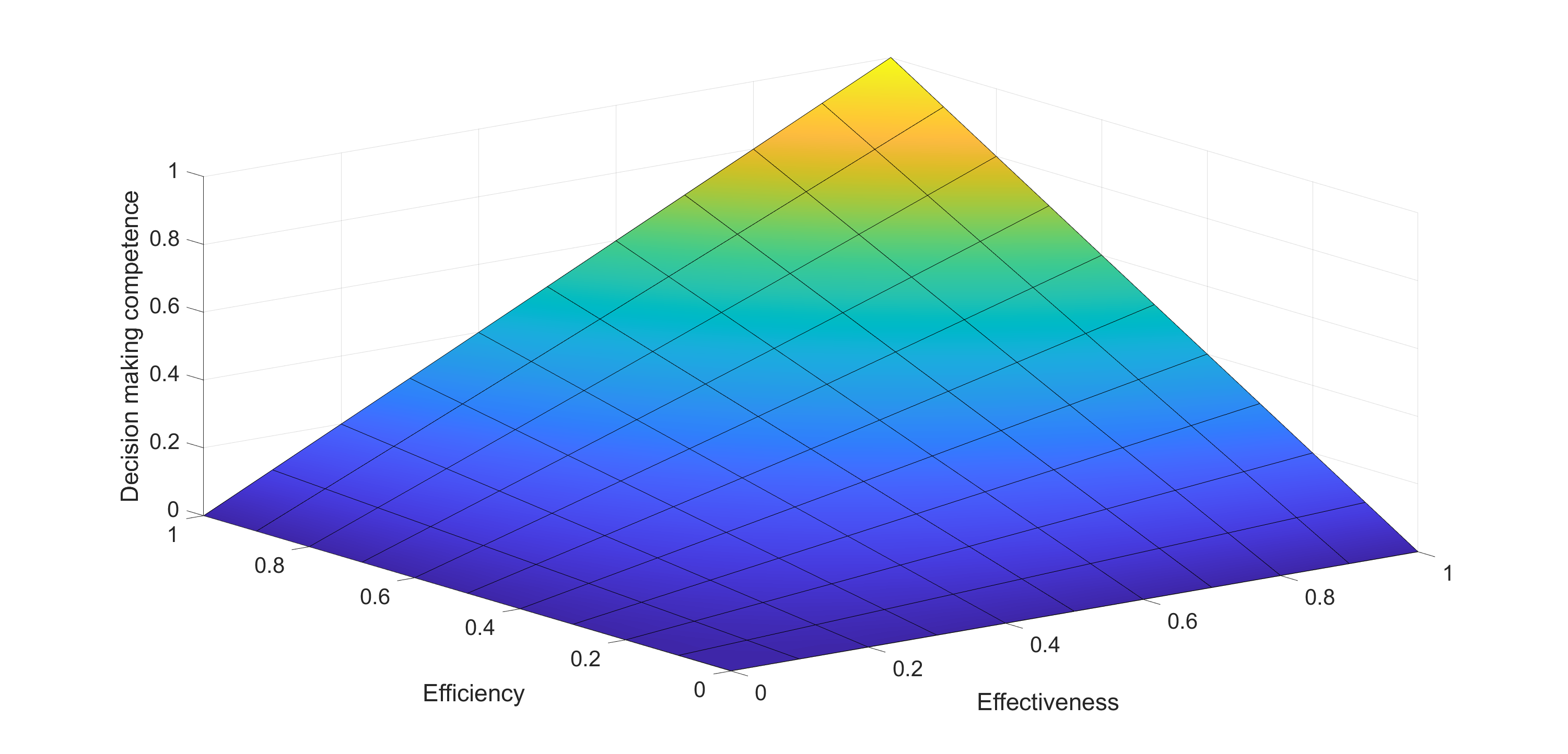}
	\caption{High intelligence in terms of decision making competence requires a combination of effectiveness and efficiency.}
	\label{fig:cPlot}
\end{figure}

Even if $c$ is hard to estimate and depends on the observer constraints $G$ and $\mathbb{S}$, the theoretical measure gives relevant implications for understanding what agent mechanisms are required for intelligent behavior.
In general, the decision making competence $c$ (Eq.~\ref{eq:int_eff}) quantifies the agent's general capability to make correct decisions weighted by spent resources (Fig.~\ref{fig:cPlot}).
Thus, an agent is only judged as intelligence if it has both the capability to make the right choices (effectiveness) and does so with acceptable effort (efficiency), e.g., in terms of reaction time.
An agent acting fast but wrong most of the time is not considered very intelligent.
In turn, an agent acting right but very slow most of the time is also not considered very intelligent.
Eq.~\ref{eq:int_eff} assumes that effectiveness and efficiency are of equal importance.
Thus, an agent acting fast but wrong most of the time may be judged as similar intelligent compared to an agent acting slow but right most of the time.
Effectiveness and efficiency are seen as interrelated; the higher the required reaction time, the less information processing capacity is needed to come up with the right solution.
Only when an agent acts both fast and right most of the time it demonstrates a higher form of intelligence according to our assumptions.

The implications of the quantification can be analyzed in more detail.
The probability of an agent to demonstrate a high effectiveness $P(g|\mathbb{S})$ (Eq.~\ref{eq:int_effective}) is driven both by the observer constraints and by the agent mechanism.
In general, in terms of observer constraints a large situation space $|\mathbb{S}|$ will make it harder for the agent to behave so that $P(g|\mathbb{S})$ is high, while in turn a larger goal set $|G|$ has the opposite effect and makes it simpler for the agent~\cite[]{Ashby1961}.
If the situation space is non-complex, because of a low size and few linear dependencies between system variables, there may exist a set of heuristic rules for the $DM$ leading to optimal performance.
In turn, if the situation space is complex, because of a large size and non-linear dependencies between system variables, an optimal agent also needs to represent a large set of different transformation rules in its $DM$ in order to model the systems dynamics~\cite[]{Conant1970a}.
Taken together, a large complex situation space with a small goal set requires a higher form of decision making competence than a small non-complex situation space with a large goal set.

Given the observer constraints, designing intelligence comes down to selecting appropriate sensors, the decision making mechanism, and effectors, and their combination into an agent architecture~\cite[]{Russell2016}.
For non-complex demands, it can be feasible to simply represent a rather full set of appropriate rules in the $DM$.
However, for complex demands, alternative approaches are required because an agent body sets an upper limit to the amount of information it can process at a time.
There are three general approaches to cope with complex demands.
First, one can aim to reduce the agent's confronted situation space by specializing in an ecological niche situation subset~\cite[]{Hasbach2021b}.
Second, an agent can attempt to continuously fit its rules to situation subsets, i.e., adapt to the current situation subset~\cite[]{Ashby1960}.
In this case, adaptation takes the form of adaptation over life cycles, i.e., evolution and design, and during life cycles, i.e., learning~\cite[]{Hasbach2021b}.
Third, an agent can attempt to extend or augment its bodily limitations, such as via tool-use and agents working together~\cite[]{Hutchins1995}.
All three strategies will be of importance for the later discussion of integrating human and \ac{ai} capabilities.

Finally, two additional strategies for increasing $c$ from Eq.~\ref{eq:int_eff} can also be derived in terms of spent resources.
First, heuristics and prepared strategies, if possible and appropriate, with low resource demands increase $c$ compared to coming up with ingenious but resource intensive strategies~\cite[]{Rasmussen1983}.
Second, projecting what could happen in the future can prevent the agent from wasting actions that may not be required, while anticipation is also relevant for decision making competence in general.

It must be noted that the above discussed implications serve only as an introduction and are by no means complete.
There are more detailed factors of the agent-world couple influencing $c$.
For example, the following list is provided by~\cite{Russell2016}, suggesting that is matters whether the task...
\begin{itemize}
	\item ...is observable or not
	\item ...deterministic or stochastic
	\item ...episodic or sequential
	\item ...static or dynamic
	\item ...time-tolerant or time-sensitive
	\item ...discrete or continuous
	\item ...known or unknown 
	\item ...single-agent or multi-agent
	\item ...competitive or cooperative.
\end{itemize}
Each selected combination changes the demand on the agent.

\subsubsection{Intelligent Agent Architecture}
\label{subsub:int_ag_ar}
\begin{figure}
	\centering
	\includegraphics[width=0.8\linewidth]{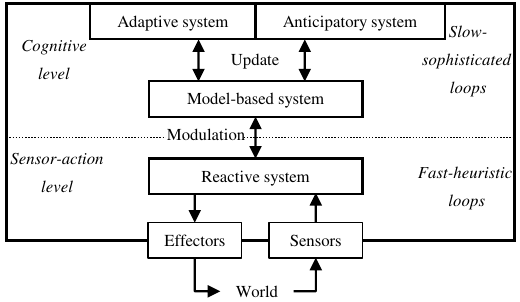}
	\caption{Blueprint of an intelligent architecture with fast reactive loops at the low level being modulated by slower but more capable cognitive loops. 
		Adapted from ~\cite{Rockbach}.}
	\label{fig:architecture}
\end{figure}

The desired properties are combined into an abstract intelligent agent architecture blueprint (Fig.~\ref{fig:architecture}).
Similar agent blueprints are discussed in both \ac{ai}~\cite[]{Russell2016} and cognitive science~\cite[]{Rasmussen1983}.
For small and non-complex regions of the situation space, fast-heuristic rules can be reactively applied in a stimulus-response fashion.
We assume that some rules are fixed in the agents, such as reflexes, while others can be updated, such as learned heuristics. 
However, for large or complex situation spaces, the cognitive system modulates the behavior of its low-level fast-heuristic arsenal, e.g., by activating and parameterizing appropriate agent subsystems~\cite[]{Minsky1988}.
This requires a representation of the problem in terms of a model, which is resource intensive to attain, however allows the agent to deal with complex problems.
Stacked on this model-based capabilities are the adaptive and anticipatory system.
The adaptive system maintains the problem representations during the agent's life, therefore continuously checks whether the model fits to the current problem and updates it if necessary. 
Finally, the anticipatory subsystem allows the agent to project to future states.

The agent architecture in Fig.~\ref{fig:architecture} shows different levels of mechanisms for decision making competence.
Easier tasks may only require fast reactive mechanisms, while harder tasks can only be solved by the slower cognitive system stacked on top.
The agent engineer must analyze the agent's task for difficulty and resource-sensitivity, and choose an appropriate mechanism level for maximizing $c$.

\section{Joint Hybrid Intelligence}
\label{sec:hi}

\subsection{Definition of Joint Hybrid Intelligence}
Given the above limitations of an embodied agent to deal with complex problems, the objective is to join human intelligence with \ac{ai} for improving on their relative decision making competencies.
\acf{si} refers to the relative competence augmentation of two or more heterogeneous agents\footnote{This includes tools only if the tool is an \ac{ai}, i.e. an artificial agent according to Def.~\ref{def:agent}.} into one joint agent demonstrating superior competence compared to the isolated individuals'~\cite[]{Hollnagel2005,Rockbach}.

On the basis of $G$ and $\mathbb{S}$, the decision making competences of individual (``internal'') agents are estimated, $c_{nat}$ for the human's intelligence and $c_{arti}$ for the \ac{ai}, and joined via an integration design in such a way that the joint intelligence $c_{joint}$ is superior to the individual's acting isolated. 
Thus, we assume that each internal agent possesses valuable decision making competencies that should be combined, i.e., we search for an integration design\footnote{With the simplest integration being human and \ac{ai} acting in parallel without interacting.} such that the joint competence $c_{joint}$ satisfies $c_{joint} > c_{nat} \vee c_{arti}$.
If $c_{joint} \leq c_{nat} \vee c_{arti}$, the design is considered disjoint and control may fully remain with either human or \ac{ai}\footnote{In some cases, $c_{joint} > c_{nat} \vee c_{arti}$ exists, however control responsibility should fully remain with the human because of ethical, legal, or organizational concerns, or fully with the \ac{ai} because human participation is not possible or desired.}.

\begin{definition}[Joint Hybrid Intelligence]

	\acf{si} refers to optimizing the joint decision making competence $c_{joint}$ by finding an integration design for heterogeneous internal agents given $G$ and $\mathbb{S}$.
	If the integration satisfies $c_{joint} > c_{nat} \vee c_{arti}$, the resulting system is called a joint (hybrid) agent.
	The practice of integrating heterogeneous internal agent aspects into a whole is referred to as \acf{jae}.
\end{definition}

\begin{figure}
	\centering
	\includegraphics[width=1\linewidth]{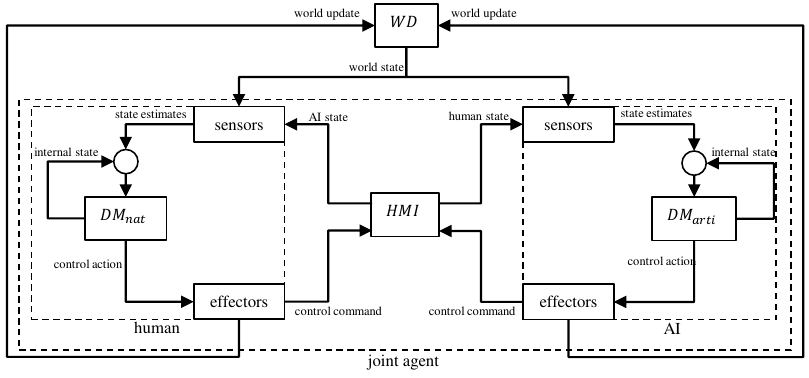}
	\caption{Human and \ac{ai} are integrated by coupling their individual decision matrices via a \acf{hmi}.}
	\label{fig:joint_agent}
\end{figure}

While in this work internal agents are humans and \ac{ai}, the definition can also be applied to other internal agent types, such as hybrid societies~\cite[]{Hamann2016}.
A joint agent provides a joint decision matrix $DM_{joint}$, it being a function of the integration design $DM_{nat} \cup DM_{arti}$.
Fig.~\ref{fig:joint_agent} shows a visualization of a simple joint agent with one human and one \ac{ai}.
Human and \ac{ai} interact with each other through a \acf{hmi}~\cite[]{Bennett2011}.
They also individually or both sense the world and act in it.
In the following, a general framework is discussed for joining human and \ac{ai}.

\subsection{Designing Joint Hybrid Intelligence}

\subsubsection{Triad of Joint Hybrid Intelligence}

\begin{figure}
	\centering
	\includegraphics[width=0.9\linewidth]{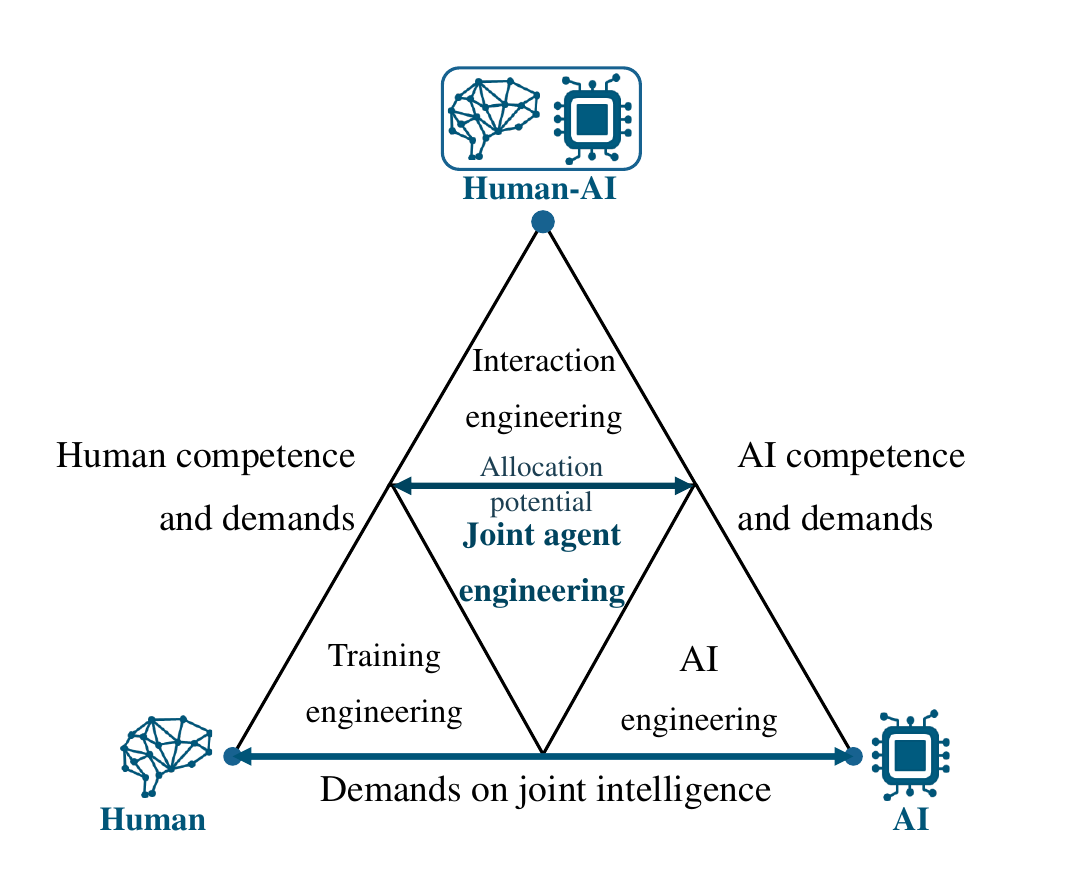}
	\caption{The triad of joint hybrid intelligence with its analysis space (outside of triad) and the embedded design subspaces (inside triads). 
		Joint agent engineering at the center integrates the different design subspaces.}
	\label{fig:triad}
\end{figure}

A visual description for designing \ac{si} is depicted by the \ac{si} triad in Fig.~\ref{fig:triad}.
The outside of the triad represents the analysis space, the inner smaller triads the design space.
The sequence of designing \ac{si} goes from bottom to top, while the center constitutes \ac{jae} with the goal of integrating the different design subspaces.

The analysis space is constituted of the following steps:
\begin{enumerate}
	\item Analyze the joint agent for demands on its intelligence
	\item Determine the individual allocation potential by comparing relative $c_{nat}$ and $c_{arti}$
\end{enumerate}

The design space is made of
\begin{enumerate}
	\item Joint agent engineering: Designing integrative patterns to bring the different design subspaces together
	\item Training engineering: Designing human training procedures for optimizing the human's $DM_{nat}$
	\item \ac{ai} engineering: Designing the \ac{ai} for optimizing the AI's $DM_{arti}$
	\item Interaction engineering: Designing the \ac{hmi} for optimizing the joint $DM_{joint}$ 
\end{enumerate}

Finally, the design is evaluated and iterated by
\begin{enumerate}
	\item Evaluating the joint agent intelligence
	\item Evaluating internal variables of the joint agent
	\item Iterate the joint agent design
\end{enumerate}

In this work, the analysis space, joint agent engineering, and the evaluation will be discussed.
For details about the specific design subspaces the interested reader is referred to \cite{Ericsson2018} for operator training, to \cite{Russell2016} for \ac{ai} engineering, and to \cite{Bennett2011} for interaction design.

\subsubsection{Demands on Joint Agent}
\label{subsub:demands}

Analyzing the joint agent demands follows the logic of Fig.~\ref{fig:intelligence}.
First, the joint agent is analyzed in terms of the observer constraints.
Here, the system analyst defines the purpose of the system, preferably in a manner so that the purpose can be formalized into a performance metric.
Next, general system dynamics for the agent-world couple are modeled in the context of the system's purpose.
Variables can be external, i.e., variables of the world external to the joint agent, or internal, i.e., variables describing the joint agent.
Another important classification is whether variables can be influenced or not.
The objective is to understand what situational context could impact performance, and to get an overview how.

A range of modeling methods can be used to model and understand the agent-world couple.
The basic technique is to develop a qualitative graph with interlinked variables.
For example, the abstraction hierarchy as part of cognitive work analysis models relevant variables from abstract to physical in the context of agent purpose~\cite[]{Rasmussen1994}.
More generally, system dynamics simulations provide a coupled set of variables and their resulting behavior~\cite[]{Forrester1971a}.
Fuzzy cognitive mapping is a technique that can also be used for simulations by providing an interlinked network of variables and is especially useful for capturing expert knowledge~\cite[]{Ozesmi2004}.  
In general, the modeler may use any qualitative or quantitative technique that advances the understanding of the agent-world couple in the context of the agent purpose.

Two levels of systems understanding can be given.
First, the analyst should aim to capture general system behaviors to get an overview of relevant system states.
Second, the utility of system states in the context of the agent purpose should be evaluated.
For example, controllable world states leading to a large probability of goal attraction are important variables to influence by effectors, since it makes the task easier for the agent.

After one has defined and analyzed the observer constraints, the required decision making competence of the joint agent is examined (Fig.~\ref{fig:intelligence}).
As defined in Sec.~\ref{subsubsec:agent_theory}, agent competence is a function of the sense-decide-act loop.
Based on the above analysis of the observer constraints, relevant variables need to be tracked by sensors and acted upon by effectors.
Here, the analyst can derive a list of requirements for what sort of information the joint agent must have access to.

The heart of the decision making competence is the decision matrix.
Here, the required information processing is helpful for understanding the demands on the decision matrix.
Task models, which are often qualitative in nature, group information processing into chunks with required input and provided output and combines them into a flow model~\cite[]{Crandall2006a}.
Following the sense-decide-act loop, a task model provides a more detailed account on the information processing demands on the joint agent.
 
The task model can then be analyzed for difficulty.
For example, classification can take the form of evaluating the required competence levels and time-sensitivity as shown in Fig.~\ref{fig:architecture}.
Another classification category can be in terms of cognitive processing with categories coordination, problem detection, sense-making, naturalistic decision making, adaptation, and planning~\cite[]{Crandall2006a}.
If the detailed task model is taken as a whole, this provides an overview of the required joint intelligence.

\subsubsection{Estimating relative Agent Intelligence}

Given the analyzed system model and task model, the individual decision making competence $c_{nat}$ and $c_{arti}$ are compared.
Such a relative comparison can be done for all tasks taken together, or for individual tasks, resulting in a task allocation potential~\cite[]{Rockbach2022b}.
The task allocation potential describes the human-\ac{ai} tension field on the basis of the relative individual competences of human and \ac{ai}, which is not to be confused with actually allocating the tasks based on an integrated view to optimize $c_{joint}$ as will be discussed in the next section~\cite[]{Dekker2002a}.
In general, task allocation levels take the form of the categories manual (only human performs task), automated (only \ac{ai} performers task), or shared (both perform some aspects of the task).
Such levels are referred to as the \ac{loa}, and often more detailed levels are used~\cite[]{Sheridan1992}.

In practice, quantitative estimations of relative $c$ are almost never feasible, given the large situation space.
However, qualitative experience exists that provide a heuristic whether the human or automation performs better for a particular task. 
For example, the classical ``Men Are Better At - Machines Are Better At'' (MABA-MABA) list provides a rough estimate what internal agent would perform better depending on the task~\cite[]{Dekker2002a}.
However, the relative internal agent intelligence must not only take the situational context into account, but also the human's experience and the sort of algorithm being used.
Therefore, defining the task allocation potential constitutes more of a design art than an engineering technique~\cite[]{Bennett2011}.

\subsubsection{Joint Agent Engineering}

At the center of the \ac{si} design space lies \ac{jae}.
The objective of \ac{jae} is to integrate the internal decision making competences in a way that the joint decision making competence is optimized based on the analyzed demands on joint agent intelligence and the task allocation potential drawn from individual competences.
In order to maximize $c_{joint}$, the joint agent's body and decision making mechanism should be able to cope with the task demands while spending as few resources as possible.
The joint agent must be scaled with sufficiently internal agents to cope with the task demands.
At the same time, internal agents should not be added blindly, since their interactions can increase internal complexity and thereby waste resources~\cite[]{Hamann2021,Woods2006a}.

\begin{figure}
	\centering
	\includegraphics[width=1\linewidth]{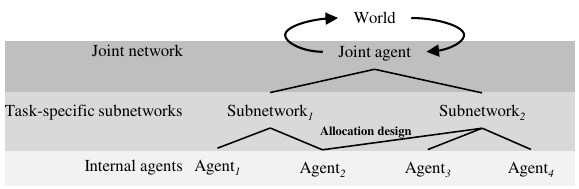}
	\caption{A joint agent is a distributed network of interacting internal agents forming into computational task-specific subnetworks.
		The allocation example shows that internal agents can contribute to multiple task-specific subnetworks.
		Adapted from~\cite{Rockbach}.}
	\label{fig:joint_agent_dis}
\end{figure}

A joint agent is a distributed network of interacting internal agents (Fig.~\ref{fig:joint_agent}) forming computational task-specific subnetworks~\cite[]{Hutchins1995}.
Internal agents can be exclusively attributed to a task corresponding to a manual or automated \ac{loa}, share the allocation with other agents (shared \ac{loa}), as well as participate in multiple tasks (Fig.~\ref{fig:joint_agent_dis}).
The allocation graph of internal agents to tasks is called the allocation design.
While today's sensors and effectors are often automated, their operations are still supervised by humans, meaning that normally both \ac{ai} and human are allocated to sensor and effector tasks.
The brain of the joint agent is modeled as the $DM_{joint}$.
Based on the task models and the associated relative allocation potential, different allocation designs can be developed.
The task allocation design heavily influences the surrounding design subspaces of the \ac{si} triad and is therefore seen as the foundation of \ac{jae}.
However, as noted above, task allocation is still more art than quantitative engineering, since coupling heterogeneous complex agents leads to unpredictability in the joint agent dynamics~\cite[]{Woods2006a}.
Different allocation designs are normally developed and qualitatively compared, e.g., via consulting system users and experts, or prototypes are implemented to quantitatively evaluate resulting joint agent competences (Sec.~\ref{subsubsec:eval}).  

To summarize, the general design problem of \ac{si} is complex, since one can come up with many possible design solutions per design dimension, resulting in scattered design facets over the actually interdependent design subspaces (Fig.~\ref{fig:triad}).
Therefore, it is helpful to come up with overreaching design heuristics, called design patterns~\cite[]{Alexander1977,Woods2006a}, serving as blueprints by providing a unified idea how the individual design parts could come together.

\begin{definition}[Joint Agent Pattern]
	A joint agent pattern is a unifying blueprint for logically constraining all design subspaces of the \ac{si} triad in order to reduce design complexity.
	In line with the objective of \ac{jae}, the patterns are aimed at maximizing the joint agent competence $c_{joint}$ by integrating different design facets into a whole. 
\end{definition}
\ac{jae}'s central objective is to find such unifying patterns and apply them in real-world contexts~\cite[]{Woods2006a}.
Two examples of joint agent patterns are discussed in the following, first for task allocation in supervisory control, and second for approaches to \ac{ai} presentation.

\begin{figure}
	\centering
	\includegraphics[width=0.5\linewidth]{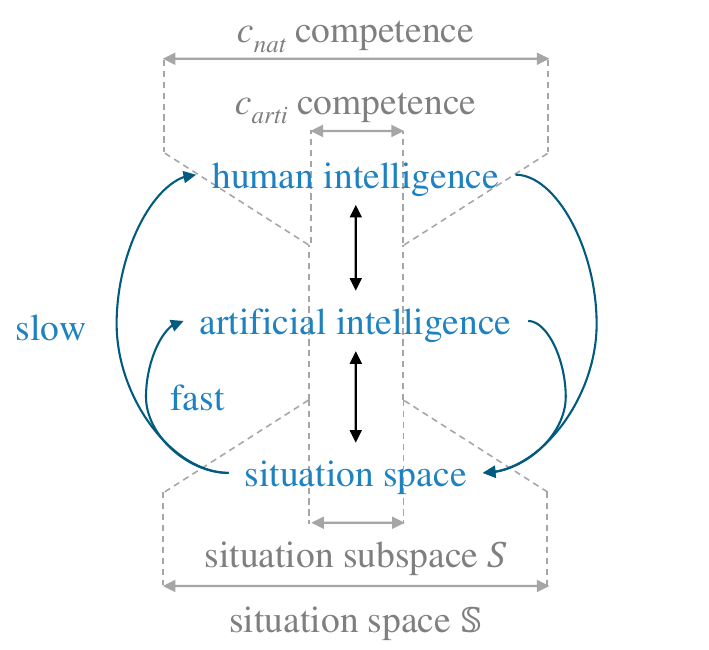}
	\caption{Joint agent architecture for supervisory control showing \ac{ai} being deployed by the human to narrow subregions of the situation space where they are allowed to act semi-autonomously.
	Adapted from \cite{Hasbach2021b}.}
	\label{fig:ja_loops}
\end{figure}

\textbf{Pattern for supervisory control.} 
Even if tasks were allocated to the \ac{ai}, human operators are often made responsible for supervising the \ac{ai}~\cite[]{Bainbridge1983,Sheridan1992}.
This can be a severe design dilemma, since the aim of automation is to reduce the human's workload. 
At the same time, however, the workload is increased by the additional task of \ac{ai} supervision.
A general joint agent architecture for automation supervision, based on the previously discussed intelligent agent architecture (Fig.~\ref{fig:architecture}), is shown in Fig.~\ref{fig:ja_loops}.
Today's \ac{ai} systems still constitute ``weak intelligence'', referring to competencies covering subregions of the situation space $S$, in comparison to the flexibility of humans' ``strong intelligence'', referring to broad situation spaces $\mathbb{S}$. 
In other words, \ac{ai} surpasses human competence in specialized tasks, i.e., for $S$, while humans surpass \ac{ai} when evaluated in $\mathbb{S}$.
Therefore, \ac{ai} is deployed for narrow subregions of the situation space for which it is specialized in, while the human's task is to check whether $c_{arti}$ fits to the current subspace and parameterize the \ac{ai} accordingly.
In this way, the different response times of humans and \ac{ai} are aligned with their competences, and supervision does not become a performance bottleneck~\cite[]{Hasbach2021b}.
This type of architecture is also known as human-on-the-loop design~\cite[]{Sheridan1992}.

\begin{figure}
	\centering
	\includegraphics[width=0.9\linewidth]{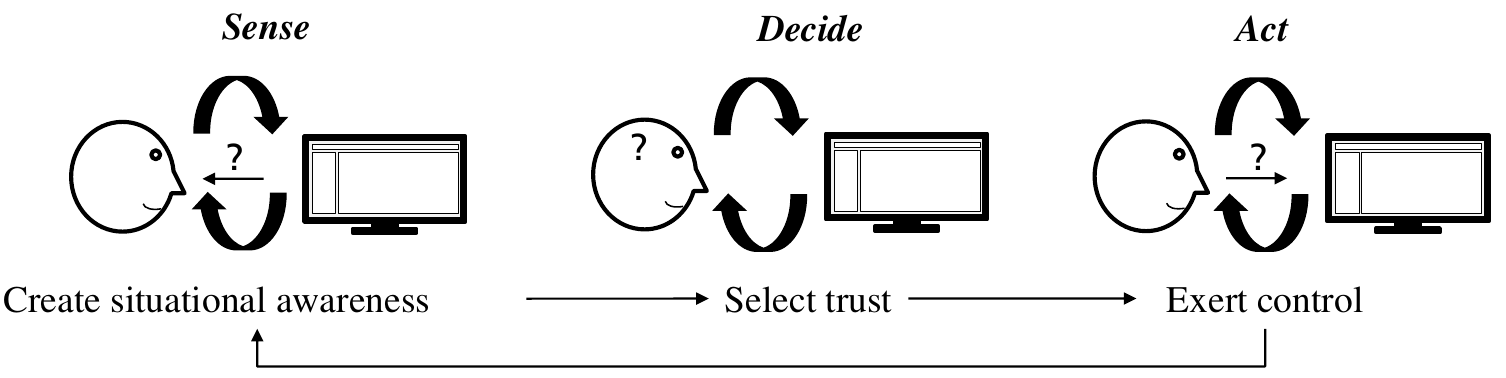}
	\caption{Sensor-decide-act steps for human supervision over \ac{ai}.
	Adapted from~\cite{Rockbach2022b}.}
	\label{fig:supervision}
\end{figure}

The design subspaces human training, \ac{ai} engineering, and interaction design are seen in light of this joint agent pattern.
For example, in order to supervise $c_{arti}$ and parameterize the \ac{ai} accordingly, the human supervisor should be experienced in implementing the following steps of the supervisory sense-decide-act loop (Fig.~\ref{fig:supervision}).
First, the human supervisor needs to understand the current state of the \ac{ai} and the world~\cite[]{Endsley1995a}, which is influenced by the transparency of \ac{ai}~\cite[]{BarredoArrieta2020} and its presentation via the \ac{hmi}.
Then, the human supervisor is required to check whether the \ac{ai} actually fits the situation subspace, it being referred to as trust calibration~\cite[]{Lee2004}.
Optimally, the operator trusts the \ac{ai} if the fit is aligned and mistrusts it if the fit is not aligned.
However, operators can also misjudge the current \ac{ai} competence.
Misuse refers to regions of overtrust where the \ac{ai} is deployed in situation subspaces with poor performance.
In turn, in disuse operators tend to deactivate the \ac{ai} even if its activation would actually increase $c_{joint}$.

\textbf{Pattern for \ac{ai} presentation}. 
Another example of a joint agent pattern are blueprints for how an \ac{ai} can be presented to the human operator.
First, an \ac{ai} can be presented as a tool to the human\footnote{Note that an \ac{ai} agent according to Def.\ref{def:agent} can still be presented as a tool to the human.}.
In the tool pattern, the main responsibility and intelligence remains with the human who uses the properties of the tools as seen fit.
Depending on the complexity of the tool, the human must be trained in its specific use, and the \ac{hmi} should provide intuition on how to control it in detail.
In contrast, the team member pattern presents the \ac{ai} as a sophisticated agent~\cite[]{ONeill2020a}.
More decision making is allocated to the \ac{ai} and the human may judge it as an intelligent system itself.
However, there is the risk of undertrust if the \ac{ai} does not hold up to the expectations or overtrust if the \ac{hmi} presents it as too intelligent.
Less training is required if the \ac{ai} demonstrates sophisticated intelligence and the \ac{hmi} design is aligned with this by communicating high level information.
Finally, the cyborg pattern treats the \ac{ai} as part of the human body and is exemplified in Sec.~\ref{sec:hsi}.

\subsubsection{Evaluation and Iteration}
\label{subsubsec:eval}

Given the unpredictability of the actual joint agent behavior, an incomplete system model being its main source, it is essential to implement a prototype and evaluate the joint agent in terms of $c_{joint}$ given ranges of situational contexts and joint agent variables.

While the specific internal variables depend on the use case and are provided by the abstraction hierarchy discussed in Sec.~\ref{subsub:demands}, supervisory control provides an important example of internal dynamics influencing $c_{joint}$.
From a human perspective, the system design should be evaluated for its tendency to equip the human operator with situational awareness over the \ac{ai}, enable trust calibration, as well as control possibilities.
Moreover, different algorithms and transparency approaches~\cite[]{Adadi2018} can be checked for their potential to fit into the joint agent as well as what level of supervisory competence is required from the human operator to use the \ac{ai} efficiently.
Finally, the joint agent can be tested for ranges of the situation spaces with different difficulties, \ac{ai} competences, and different levels of operator training, to identify joint performance regions of ``brittleness''~\cite[]{Woods2006a}, i.e., regions where $c_{joint}$ suddenly breaks down.
Such analysis helps to understand the system dynamics and to update the selected subdesigns of the \ac{si} triad. 

Often, it is not feasible to implement a range of complete prototypes to compare between different designs.
It can, however, be feasible to design low-fidelity prototypes.
For example, the wizard-of-oz technique only simulates \ac{ai} behavior that is controlled by a hidden experimenter~\cite[]{Maulsby1993}. 
In addition, ``staged world experiments'' increase the ecological validity of experimental lab settings, e.g., by providing a realistic contextual scenario, and have been discussed as a helpful practical trade-off between controllable experiments and ecological complexity~\cite[]{Woods2006a}.

\section{Human Control over Robot Swarms}
\label{sec:hsi}

\subsection{Swarm Supervision as Design Problem}
Swarm robotics envisions robot collectives in which rather simple robots interact only with local neighbors without reference to global information~\cite[]{Dorigo2021}.
This decentralized architecture makes robot swarms both scalable and robust to single robot failures.
However, in order to deploy robot swarms to real-world problems, human operators must be in control of these decentralized systems~\cite[]{Kolling2016}. 
Such \acf{hsi} is a particular case of \ac{si}.

\ac{hsi} comes with two special design challenges.
First, integrating a human supervisor into the robot swarm introduces centralization into the robot swarm that was deliberately designed in a decentralized way, since the human must be able to influence the whole swarm and be provided with state estimations from the swarm.
This is referred to as centralized-decentralized trade-off~\cite[]{Hasbach2022}.
Second, supervisory control methods must take the swarm's inherent scalability into account, meaning that the control complexity should be invariant to the swarm size.
For example, a human operator could control a leader robot which influences the rest of the swarm via local interactions~\cite[]{Kolling2016}. 

Work in \ac{hsi} has mainly focused on the aspects of swarm engineering and interaction design.
However, it was discussed here that the design space is a function of the integrated aspects of the \ac{si} triad.
Therefore, we summarize recent work on a joint agent pattern for \ac{hsi} based on the cyborg pattern~\cite[]{Clark2003}, in the following referred to as ``extended swarming''.  

\subsection{Swarms as Controllable Embodied Extensions}
In the joint agent pattern ``extended swarming'', the swarm is designed as an embodied extension of the human operator~\cite[]{Hasbach2022}.
The pattern is treated as a scientific theory that must generate testable design solutions for the \ac{si} design subspace given the hypothesis of good joint agent performance. 
From the human's perspective, the swarm becomes a sensation-action extension of oneself.
A human who is extended by a self-organizing swarm is referred to as swarm-amplified human, since the extension of the human's low-level sensation-action range is hypothesized to amplify the human's high-level natural cognitive capacities by augmentation.
Given this cyborg pattern, the ultimate design goal is to induce the ``sense of ownership''~\cite[]{Braun2018} towards the swarm into the human operator, but the pattern is also useful without such perceived embodiment~\cite[]{Rockbach2022a}. 
In turn, from the swarm's perspective the human becomes a special swarm agent while robots interact only locally with the human.
These two perspectives are required to balance the centralized-decentralized trade-off.
In the following, the different design subspaces are shortly addressed in the light of extended swarming.

\subsubsection{Operator Training}

Little work has considered operator training in \ac{hsi} although the human's expertise of a technical system modulates joint agent dynamics~\cite[]{Ericsson2018}.
If the swarm is treated as an extension of the human, the training follows the learning of body control in humans via trial-and-error~\cite[]{Hasbach2022}.
The joint human-swarm agent is exposed to a range of situations with different degrees of complexity~\cite[]{Hasbach2020}.
The goal is to develop an intuitive mental model of swarm control, similar to the feeling we have when controlling our own bodies~\cite[]{Rockbach2022a}.

\subsubsection{Swarm Engineering}

\begin{figure}
	\centering
	\includegraphics[width=0.7\linewidth]{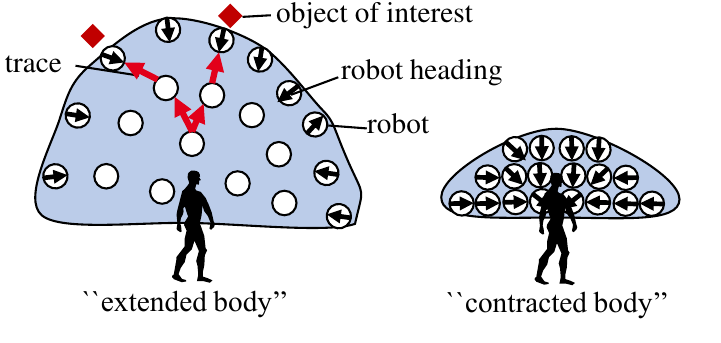}
	\caption{Example of extending body postures implemented by a robot swarm with an embedded fusion network.
	The pose is the observed collective state of the robot swarm, shown by the blue region.
	The distributed fusion network points towards sensed states (red arrows), such as objects of interest, and can be observed and used by the human as a movement gradient.}
	\label{fig:hsi_posture}
\end{figure}

Swarms must both provide the possibility to be controlled by the human and provide fused estimates about the swarm state~\cite[]{Kolling2016}.
The swarm is designed as a self-organizational network with the human agent at the center of it, while only minimally reducing the swarm's decentralized nature.  
In terms of controllability, the swarm self-organizes into ``extending body postures'' based on human input and environmental states~\cite[]{Rockbach2023}.
For example, a swarm state clustering at the human corresponds to a ``contracted body state'', while a dispersed swarm is an ``extended body state'' (Fig.~\ref{fig:hsi_posture}).

Information fusion is implemented in a decentralized manner by the swarm itself.
The swarm self-organizes into a decentralized fusion network with the human operator at the center of it based on the environmental state~\cite[]{Rockbach2023a}.
The fusion network can be thought of as a spider web providing a 360° input of the sensed environment, thereby extending the human's sensation range (Fig.~\ref{fig:hsi_posture}).
In addition, the spider web can also be used to allocate own agents to particular areas, e.g., the human can follow traces of the spider web towards objects of interest.

\subsubsection{Interface Design}

In extended swarming, the swarm is treated as an extension of the human's sensation-action loop with the world.
Thus, the swarm itself becomes the interface between the human and the world such as discussed in terms of decentralized information fusion above.
From the human's control perspective, it is hypothesized that classifying human body states, such as body posture or physiological states, and utilizing them to automatically influence the swarm posture is beneficial for overall performance~\cite[]{Hasbach2022, Rockbach2022a}.
This is derived from the general experience with intuitive body control, which is mainly subconscious in everyday life.
In addition, the design must feature the possibility for deliberate control, e.g., by controlling an extended limb formed by the swarm~\cite[]{Rockbach2023} with one's own arm, as one can deliberately control detailed aspects of the body when one desires to do so. 

The feedback provided to the human would take the form of low-level sensation stimulation such as an increase in a visual or tactile stimulus when facing a state of interest sensed by the swarm extension~\cite[]{Hasbach2022, Rockbach2022a}.
This also reflects daily-life embodied phenomenology, e.g., when coming closer to a fire increases the feeling of body warmth.
Based on such low-level stimulation, the brain's high-level operations fuse and embed the low-level input into a holistic percept of the current agent-body state given an adapted human's model of body and world.

\section{Conclusion}
\label{sec:con}

This chapter discussed an approach for joining the heterogeneous capabilities of humans and \ac{ai} into a joint hybrid agent.
Intelligence was formalized as decision making competence to be applicable to both humans and \ac{ai}.
This formalization was analysed for designing intelligence, including the augmentation of one agent by another agent, providing the theoretical basis for \acf{si}. 
\ac{si} was presented as the optimization of joint decision making competence by integrating the design subspaces operator training, \ac{ai} engineering, and interface design via the development of overreaching joint agent patterns.
Finally, ``extended swarming'' provided an example of a joint agent pattern for human-swarm interaction.

While the integration of humans and \ac{ai} has a long history, it is important to explicitly develop a framework which brings the different disciplines and their terminologies together.
\ac{si} provides a common basis for the multiple scientific and engineering branches concerned with human-\ac{ai} integration, and therefore facilitates the integration of human and \ac{ai} into intelligent joint decision making agents.

\bibliography{bib3}

@book{Rasmussen1994,
author = {Rasmussen, Jens and Pejtersen, Annelise Mark and Goodstein, Len P},
publisher = {Wiley},
title = {{Cognitive systems engineering}},
year = {1994}
}

@inproceedings{Rockbach2023,
author = {Rockbach, Jonas D},
booktitle = {2023 IEEE International Conference on Systems, Man, and Cybernetics (SMC)},
title = {{Enhancing Human Self-Regulation with Controllable Robot Swarms Acting as Extended Bodies}},
year = {2023}
}

@book{Brooks1999,
author = {Brooks, Rodney Allen},
isbn = {0262522632},
publisher = {MIT press},
title = {{Cambrian intelligence: The early history of the new AI}},
year = {1999}
}

@book{Woods2006a,
author = {Woods, David D. and Hollnagel, Erik},
booktitle = {Joint Cognitive Systems: Patterns in Cognitive Systems Engineering},
doi = {10.1080/00140130701223774},
isbn = {9781420005684},
issn = {0014-0139},
publisher = {CRC Press},
title = {{Joint cognitive systems: Patterns in cognitive systems engineering}},
year = {2006}
}

@article{Rosenblueth1943,
author = {Rosenblueth, Arturo and Wiener, Norbert and Bigelow, Julian},
issn = {0031-8248},
journal = {Philosophy of science},
number = {1},
pages = {18--24},
publisher = {Cambridge University Press},
title = {{Behavior, purpose and teleology}},
volume = {10},
year = {1943}
}

@book{Ashby1961,
author = {Ashby, W Ross},
publisher = {Chapman {\&} Hall Ltd},
title = {{An introduction to cybernetics}},
year = {1961}
}

@book{Hutchins1995,
author = {Hutchins, Edwin},
isbn = {0262581469},
publisher = {MIT press},
title = {{Cognition in the Wild}},
year = {1995}
}

@article{Flemisch2012b,
author = {Flemisch, Frank and Heesen, Matthias and Hesse, Tobias and Kelsch, Johann and Schieben, Anna and Beller, Johannes},
issn = {1435-5566},
journal = {Cognition, Technology {\&} Work},
number = {1},
pages = {3--18},
publisher = {Springer},
title = {{Towards a dynamic balance between humans and automation: authority, ability, responsibility and control in shared and cooperative control situations}},
volume = {14},
year = {2012}
}

@article{Hamann2016,
author = {Hamann, Heiko and Khaluf, Yara and Botev, Jean and {Divband Soorati}, Mohammad and Ferrante, Eliseo and Kosak, Oliver and Montanier, Jean-Marc and Mostaghim, Sanaz and Redpath, Richard and Timmis, Jon},
issn = {2296-9144},
journal = {Frontiers in Robotics and AI},
pages = {14},
publisher = {Frontiers Media SA},
title = {{Hybrid societies: challenges and perspectives in the design of collective behavior in self-organizing systems}},
volume = {3},
year = {2016}
}

@article{Bainbridge1983,
author = {Bainbridge, Lisanne},
doi = {10.1016/0005-1098(83)90046-8},
isbn = {0005-1098},
issn = {00051098},
journal = {Automatica},
title = {{Ironies of automation}},
year = {1983}
}

@book{Downing2015,
author = {Downing, Keith L},
isbn = {0262029138},
publisher = {MIT Press},
title = {{Intelligence emerging: adaptivity and search in evolving neural systems}},
year = {2015}
}

@book{Conant1981,
author = {Conant, Roger},
isbn = {1127197703},
publisher = {Intersystems Publ.},
title = {{Mechanisms of intelligence: Ashby's writings on cybernetics}},
year = {1981}
}

@article{Rasmussen1983,
author = {Rasmussen, Jens},
issn = {0018-9472},
journal = {IEEE transactions on systems, man, and cybernetics},
number = {3},
pages = {257--266},
publisher = {IEEE},
title = {{Skills, rules, and knowledge; signals, signs, and symbols, and other distinctions in human performance models}},
year = {1983}
}

@book{Minsky1988,
author = {Minsky, Marvin},
isbn = {0671657135},
publisher = {Simon and Schuster},
title = {{The society of mind}},
year = {1988}
}

@inproceedings{Rockbach,
author = {Rockbach, Jonas D and Fuchs, Sven and Witte, Thomas E.F. and Bluhm, Luka-Franziska},
booktitle = {2022 IEEE International Conference on Human-Machine Systems (ICHMS)},
title = {{Ingredients for Hybrid Intelligence: Towards an Integrated Theory and Application}},
year = {2022}
}

@article{Hamann2021,
author = {Hamann, Heiko and Reina, Andreagiovanni},
issn = {0018-9340},
journal = {IEEE Transactions on Computers},
publisher = {IEEE},
title = {{Scalability in computing and robotics}},
year = {2021}
}

@article{Forrester1971a,
author = {Forrester, Jay W},
issn = {0040-5833},
journal = {Theory and decision},
number = {2},
pages = {109--140},
publisher = {Springer},
title = {{Counterintuitive behavior of social systems}},
volume = {2},
year = {1971}
}

@book{Bennett2011,
author = {Bennett, Kevin B and Flach, John M},
isbn = {1420064398},
publisher = {CRC Press},
title = {{Display and interface design: Subtle science, exact art}},
year = {2011}
}

@book{Hollnagel2005,
author = {Hollnagel, Erik and Woods, David D.},
doi = {10.1177/106480460701500208},
isbn = {0-8493-2821-7},
issn = {1064-8046},
pages = {241},
pmid = {13796389},
publisher = {CRC Press},
title = {{Joint Cognitive Systems. Foundations of Cognitive Systems Engineering}},
year = {2005}
}

@article{ONeill2020a,
author = {O'Neill, Thomas and McNeese, Nathan and Barron, Amy and Schelble, Beau},
doi = {10.1177/0018720820960865},
issn = {15478181},
journal = {Human Factors},
title = {{Human–Autonomy Teaming: A Review and Analysis of the Empirical Literature}},
year = {2020}
}

@article{Kolling2016,
author = {Kolling, Andreas and Walker, Phillip and Chakraborty, Nilanjan and Sycara, Katia and Lewis, Michael},
doi = {10.1109/THMS.2015.2480801},
isbn = {2013631553},
issn = {21682291},
journal = {IEEE Transactions on Human-Machine Systems},
number = {1},
pages = {9--26},
pmid = {7299280},
title = {{Human Interaction with Robot Swarms: A Survey}},
volume = {46},
year = {2016}
}

@article{Adadi2018,
author = {Adadi, Amina and Berrada, Mohammed},
doi = {10.1109/ACCESS.2018.2870052},
issn = {21693536},
journal = {IEEE Access},
pages = {52138--52160},
publisher = {IEEE},
title = {{Peeking Inside the Black-Box: A Survey on Explainable Artificial Intelligence (XAI)}},
volume = {6},
year = {2018}
}

@inproceedings{Rockbach2023a,
author = {Rockbach, Jonas D and Schlangen, Isabel and Bennewitz, Maren},
booktitle = {2023 IEEE Symposium Sensor Data Fusion and International Conference on Multisensor Fusion and Integration (SDF-MFI)},
isbn = {9798350382587},
pages = {1--6},
publisher = {IEEE},
title = {{Self-organising Distributed Sensor Fusion Networks for Hierarchical Swarm Control and Supervision}},
year = {2023}
}

@article{Dorigo2021,
author = {Dorigo, Marco and Theraulaz, Guy and Trianni, Vito},
journal = {Proceedings of the IEEE},
number = {7},
pages = {1152--1165},
title = {{Swarm Robotics: Past, Present, and Future [Point of View]}},
volume = {109},
year = {2021}
}

@book{Ashby1960,
author = {Ashby, W. Ross},
edition = {Second},
publisher = {Chapman {\&} Hall Ltd},
title = {{Design For A Brain: The origin of adaptive behaviour}},
year = {1960}
}

@book{Ericsson2018,
author = {Ericsson, K Anders and Hoffman, Robert R and Kozbelt, Aaron},
isbn = {1107137551},
publisher = {Cambridge University Press},
title = {{The Cambridge handbook of expertise and expert performance}},
year = {2018}
}

@article{Conant1970a,
author = {Conant, Roger C. and {Ross Ashby}, W.},
doi = {10.1080/00207727008920220},
issn = {14645319},
journal = {International Journal of Systems Science},
title = {{Every good regulator of a system must be a model of that system}},
year = {1970}
}

@book{Alexander1977,
author = {Alexander, Christopher},
isbn = {0199726531},
publisher = {Oxford university press},
title = {{A pattern language: towns, buildings, construction}},
year = {1977}
}

@book{Russell2016,
author = {Russell, Stuart and Norvig, Peter},
edition = {3rd},
publisher = {Pearson},
title = {{Artificial Intelligence: A Modern Approach}},
year = {2016}
}

@article{Stanney2009,
author = {Stanney, Kay M and Schmorrow, Dylan D and Johnston, Matthew and Fuchs, Sven and Jones, David and Hale, Kelly S and Ahmad, Ali and Young, Peter},
issn = {1557-234X},
journal = {Reviews of human factors and ergonomics},
number = {1},
pages = {195--224},
publisher = {SAGE Publications Sage CA: Los Angeles, CA},
title = {{Augmented cognition: An overview}},
volume = {5},
year = {2009}
}

@article{Licklider1960,
author = {Licklider, J. C R},
doi = {10.1109/THFE2.1960.4503259},
issn = {21682836},
journal = {IRE Transactions on Human Factors in Electronics},
title = {{Man-Computer Symbiosis}},
year = {1960}
}

@article{Lee2004,
author = {Lee, John D and See, Katrina A},
issn = {0018-7208},
journal = {Human factors},
number = {1},
pages = {50--80},
publisher = {SAGE Publications Sage UK: London, England},
title = {{Trust in automation: Designing for appropriate reliance}},
volume = {46},
year = {2004}
}

@article{Endsley1995a,
author = {Endsley, Mica R},
issn = {0018-7208},
journal = {Human factors},
number = {1},
pages = {32--64},
publisher = {SAGE Publications Sage CA: Los Angeles, CA},
title = {{Toward a theory of situation awareness in dynamic systems}},
volume = {37},
year = {1995}
}

@article{Braun2018,
author = {Braun, Niclas and Debener, Stefan and Spychala, Nadine and Bongartz, Edith and S{\"{o}}r{\"{o}}s, Peter and M{\"{u}}ller, Helge H.O. and Philipsen, Alexandra},
doi = {10.3389/fpsyg.2018.00535},
issn = {16641078},
journal = {Frontiers in Psychology},
number = {APR},
pages = {1--17},
title = {{The senses of agency and ownership: A review}},
volume = {9},
year = {2018}
}

@inproceedings{Hasbach2021b,
author = {Hasbach, Jonas D. and Witte, Thomas E.F.},
booktitle = {2021 IEEE International Conference on Systems, Man, and Cybernetics (SMC)},
title = {{Human-Machine Intelligence: Frigates are Intelligent Organisms}},
year = {2021}
}

@book{Sheridan1992,
author = {Sheridan, Thomas B},
isbn = {0262193167},
publisher = {MIT press},
title = {{Telerobotics, automation, and human supervisory control}},
year = {1992}
}

@article{Ozesmi2004,
author = {{\"{O}}zesmi, Uygar and {\"{O}}zesmi, Stacy L.},
doi = {10.1016/j.ecolmodel.2003.10.027},
isbn = {0304-3800},
issn = {03043800},
journal = {Ecological Modelling},
number = {1-2},
pages = {43--64},
title = {{Ecological models based on people's knowledge: A multi-step fuzzy cognitive mapping approach}},
volume = {176},
year = {2004}
}

@book{Simon1982,
author = {Simon, H},
publisher = {MIT press},
title = {{The Sciences of the Artificial}},
year = {1982}
}

@book{Wiener1961,
author = {Wiener, Norbert},
isbn = {0262355914},
publisher = {MIT press},
title = {{Cybernetics or Control and Communication in the Animal and the Machine}},
year = {1961}
}

@inproceedings{Rockbach2022a,
author = {Rockbach, Jonas D and Bennewitz, Maren},
booktitle = {IOP Conference Series: Materials Science and Engineering},
isbn = {1757-899X},
number = {1},
pages = {12015},
publisher = {IOP Publishing},
title = {{Robot swarms as embodied extensions of humans}},
volume = {1261},
year = {2022}
}

@article{Dekker2002a,
author = {Dekker, Sidney W A and Woods, David D},
issn = {1435-5558},
journal = {Cognition, Technology {\&} Work},
pages = {240--244},
publisher = {Springer},
title = {{MABA-MABA or abracadabra? Progress on human–automation co-ordination}},
volume = {4},
year = {2002}
}

@inproceedings{Maulsby1993,
author = {Maulsby, David and Greenberg, Saul and Mander, Richard},
booktitle = {Proceedings of the INTERACT'93 and CHI'93 conference on Human factors in computing systems},
pages = {277--284},
title = {{Prototyping an intelligent agent through Wizard of Oz}},
year = {1993}
}

@inproceedings{Hasbach2020,
author = {Hasbach, Jonas D and Witte, Thomas E F and Bennewitz, Maren},
booktitle = {International Conference on Human-Computer Interaction},
pages = {311--329},
publisher = {Springer},
title = {{On the Importance of Adaptive Operator Training in Human-Swarm Interaction}},
year = {2020}
}

@book{Shapiro2019,
author = {Shapiro, Lawrence},
isbn = {1351719165},
publisher = {Routledge},
title = {{Embodied cognition}},
year = {2019}
}

@article{Hasbach2022,
author = {Hasbach, Jonas D and Bennewitz, Maren},
issn = {1059-7123},
journal = {Adaptive Behavior},
number = {4},
pages = {361--386},
publisher = {SAGE Publications Sage UK: London, England},
title = {{The design of self-organizing human–swarm intelligence}},
volume = {30},
year = {2022}
}

@article{Newen2018,
author = {Newen, Albert and Gallagher, Shaun and {De Bruin}, Leon},
title = {{4E cognition: Historical roots, key concepts, and central issues}},
year = {2018}
}

@article{BarredoArrieta2020,
archivePrefix = {arXiv},
arxivId = {1910.10045},
author = {{Barredo Arrieta}, Alejandro and D{\'{i}}az-Rodr{\'{i}}guez, Natalia and {Del Ser}, Javier and Bennetot, Adrien and Tabik, Siham and Barbado, Alberto and Garcia, Salvador and Gil-Lopez, Sergio and Molina, Daniel and Benjamins, Richard and Chatila, Raja and Herrera, Francisco},
doi = {10.1016/j.inffus.2019.12.012},
eprint = {1910.10045},
issn = {15662535},
journal = {Information Fusion},
pages = {82--115},
title = {{Explainable Explainable Artificial Intelligence (XAI): Concepts, taxonomies, opportunities and challenges toward responsible AI}},
volume = {58},
year = {2020}
}

@book{Clark2003,
author = {Clark, Andy},
publisher = {Oxford university press},
title = {{Natural-Born Cyborgs: Minds, Technologies, and the Future of Human Intelligence}},
year = {2003}
}

@book{Varela1991a,
author = {Varela, Francisco J and Thompson, Evan and Rosch, Eleanor},
isbn = {0262335506},
publisher = {MIT Press},
title = {{The Embodied Mind: Cognitive Science and Human Experience}},
year = {1991}
}

@book{Maturana1970,
author = {Maturana, Humberto R and {Von Foerster}, H and Weston, P},
publisher = {University of Illinois Urbana},
title = {{Neurophysiology of cognition}},
year = {1970}
}

@book{Crandall2006a,
author = {Crandall, Beth and Klein, Gary A and Hoffman, Robert R},
isbn = {0262532816},
publisher = {Mit Press},
title = {{Working minds: A practitioner's guide to cognitive task analysis}},
year = {2006}
}

@inproceedings{Rockbach2022b,
author = {Rockbach, Jonas D and Bluhm, Luka-Franziska and Schlangen, Isabel and Over, Laura and Apfeld, Sabine and Henneke, Lukas and Wilkinghoff, Kevin},
booktitle = {2022 Sensor Data Fusion: Trends, Solutions, Applications (SDF)},
isbn = {1665486724},
pages = {1--6},
publisher = {IEEE},
title = {{Towards Human-Machine Integration for Signals Intelligence Applications}},
year = {2022}
}
\bibliographystyle{apalike}
\end{document}